\documentclass{article}
\usepackage{xcolor,colortbl}
\usepackage{color,soul}
\definecolor{Gray}{gray}{0.85}
\newcolumntype{C}{>{\centering\let\newline\\\arraybackslash\hspace{0pt}}m{0.1\textwidth}}
\newcolumntype{A}{>{\centering\let\newline\\\arraybackslash\hspace{0pt}}m{0.05\textwidth}}
\newcolumntype{V}{>{\centering\let\newline\\\arraybackslash\hspace{0pt}}m{0.4\textwidth}}
\newcolumntype{L}{>{\let\newline\\\arraybackslash\hspace{0pt}}m{0.85\textwidth}}
\newcolumntype{B}{>{\let\newline\\\arraybackslash\hspace{0pt}}m{0.82\textwidth}}
\newcolumntype{M}{>{\let\newline\\\arraybackslash\hspace{0pt}}m{0.1\textwidth}}
\usepackage{float}
\usepackage{comment}
\usepackage{placeins}
\usepackage{makecell}
\usepackage{array}
\usepackage{comment}
\usepackage[nolist]{acronym}

\usepackage{subfig}
\usepackage{amsmath}
\usepackage{enumitem}
\usepackage{longtable}
\usepackage{xcolor}
\usepackage{comment}

\usepackage{arxiv}

\usepackage[utf8]{inputenc} 
\usepackage[T1]{fontenc}    
\usepackage{hyperref}       
\usepackage{url}            
\usepackage{booktabs}       
\usepackage{amsfonts}       
\usepackage{nicefrac}       
\usepackage{microtype}      
\usepackage{lipsum}		    
\usepackage{graphicx}
\usepackage{natbib}
\usepackage{doi}
\usepackage{enumerate}
\usepackage[nolist]{acronym}
\usepackage{todonotes}
\usepackage{pdfpages}

\newcommand{\trtitle}{Participant Perceptions of a Robotic Coach Conducting Positive Psychology
Exercises: A Qualitative Analysis}
\title{\trtitle}

\author{Minja~Axelsson$^{1}$, Nikhil~Churamani$^{1}$, Atahan~Caldir$^{2}$ and Hatice~Gunes$^{1}$\\
	$^{1}$ Department of Computer Science and Technology, 
	University of Cambridge, Cambridge, UK\\
        $^{2}$ Department of Computer Science, 
        Özyeğin University, Turkey\\ 
 	\texttt{\{minja.axelsson, nikhil.churamani, hatice.gunes\}@cl.cam.ac.uk}
}

\renewcommand{\shorttitle}{Participant Perceptions of a Robotic Coach Conducting Positive Psychology
Exercises: A Qualitative Analysis}

\hypersetup{
pdftitle={\trtitle},
pdfsubject={},
pdfauthor={Minja~Axelsson, Nikhil~Churamani, Atahan~Caldir and Hatice~Gunes},
pdfkeywords={Human-Robot Interaction, Well-being, Robot Design, Positive Psychology, Personalisation, Socially Assistive Robotics},
}

\renewcommand\hl[1]{#1} 

\begin{document}
\maketitle

 \begin{abstract}
This paper presents a qualitative analysis of participants' perceptions of a robotic coach conducting Positive Psychology exercises, providing insights for the future design of robotic coaches. Participants $(n = 20)$ took part in a single-session (avg. $31\pm10$ minutes) Human-Robot Interaction study in a laboratory setting. We created the design of the robotic coach, and its \emph{affective adaptation}, based on user-centred design research and collaboration with a professional coach. We transcribed post-study participant interviews and conducted a Thematic Analysis. We discuss the results of that analysis, presenting aspects participants found particularly helpful (e.g., the robot asked the correct questions and helped them think of new positive things in their life), and what should be improved (e.g., the robot's utterance content should be more responsive). We found that participants had no clear preference for affective adaptation or no affective adaptation, which may be due to both positive and negative user perceptions being heightened in the case of adaptation. Based on our qualitative analysis, we highlight insights for the future design of robotic coaches, and areas for future investigation (e.g., examining how participants with different personality traits, or participants experiencing isolation, could benefit from an interaction with a robotic coach).
 \end{abstract}
\vspace{-5mm}

\section{Introduction}
\label{sec:introduction}

The need for mental health interventions is increasing with more individuals turning towards different well-being practices such as mindfulness~\cite{barnes2017examination,wimmer2019improving} and \acf{PP}~\cite{gable2005and}. While mindfulness provides a meditation-based tool for alleviating anxiety and depression levels~\cite{NHSMindfulness}, \ac{PP} is a branch of psychology that aims to enhance well-being by focusing particularly on the positive experiences of people and positive individual traits~\cite{gable2005and,seligman2014positive}. Such positive reflection has been shown to increase feelings of positive affect and future expectancy in individuals~\cite{peters2010manipulating}. 


With the recent COVID-19 pandemic, the general population has been severely impacted, resulting in negative mental health outcomes~\cite{talevi2020mental}. People have found it particularly difficult to seek mental health advice and treatment due to social distancing regulations imposed~\cite{Li2020Challenges}. As a result, digital forms of healthcare have been applied to assist individuals in need~\cite{moreno2020mental}. These include tele-health services via video calls, online therapies, and self-help resources through mobile and web apps. As more people become accepting of robots as mental health support partners~\cite{Orcale2020}, \acp{SAR} are being explored to provide mental health coaching and assistance~\cite{scoglio2019use, robinson2021robot}. Examples include robots aimed at improving the psychological well-being of individuals by providing mindfulness coaching~\cite{alimardani2020robot, Bodala2021Teleoperated}, instilling behavioural changes to achieve weight loss~\cite{kidd2008robots}, and administering \ac{PP} coaching~\cite{jeong2020robotic, jeong2023deploying}. 
In dearth of close access to experienced mental well-being practitioners, such \acsp{SAR} could support individuals and help prevent the deterioration of mental health to the point where medical intervention is needed, by providing accessible and timely assistance. 

\textcolor{black}{In fact, our previous work investigating the design of a robotic well-being coach together with users and professional coaches identified accessibility as a potential advantage that robotic coaches could have over human coaches \cite{axelsson2021roman}. Participants also noted that in comparison to a human coach, a robot would not get tired and could be more consistent due to lack of being affected by a daily routine \cite{axelsson2021roman}, and that its physical presence was an advantage over mobile-based well-being applications. However, a robot was also seen to have disadvantages, including technological malfunctions and unwanted robot behaviour, ``lack of human-ness'' (e.g., lack of personal history and world experience), as well as higher cost in comparison to a mobile application \cite{axelsson2021roman}. Another longitudinal study examined the use of a robot as a mindfulness coach, in comparison to a human coach~\cite{Bodala2021Teleoperated}. Both coaches received positive feedback, however the human coach was rated significantly higher than the robotic coach \cite{Bodala2021Teleoperated}. \citet{duan2021self} compared self-disclosure to a robot and in a journal, and found that people who felt stronger negative emotions benefited more from talking to the robot in comparison to the journal. Another study compared a robot and a voice agent delivering a journalling-based well-being exercise, finding that participants had higher mood improvements and levels of self-disclosure when interacting with the robot \cite{sayis2024technology}. Similarly, \citet{alves2022robot} compared a traditional workbook and a robot for supporting adolescents' mental health, finding that adolescents preferred the robot, and that it created more engagement. While a robot may not reach the level of performance of a human coach (cf. \cite{Bodala2021Teleoperated}), these studies indicate that they have advantages over other technological and traditional approaches to mental well-being exercises. This highlights the importance of investigating \acsp{SAR} as potential tools for delivering well-being practices.}

For \acsp{SAR} to be effective, they need socio-emotional adaptability that allows them to not only generalise to real-world interactions with individuals but also provide personalised interaction experiences by adapting to individual context and behaviours~\cite{Churamani2020CL4AR}. They need to be sensitive to specific user verbal and non-verbal behaviours~\cite{Churamani2020CLIFER} while learning to respond dynamically and appropriately, adhering to the context and evolution of the interaction~\cite{Dautenhahn2004Robots, Ficocelli2016Promoting}. In fact, prospective users of a robotic coach have expressed a need for a robotic coach to be adaptive to their emotional expressions, and professional coaches also agreed that this can be useful in the context of coaching \cite{axelsson2021roman}.

\hl{In this study, we create an autonomously functioning implementation of a robotic Positive Psychology (PP) coach, and conduct a user study in a controlled laboratory environment. The design of the robotic coach was based on our prior user-centred design study} \cite{axelsson2021roman}, \hl{where we generated insights for robotic coach design with users and professional coaches. In our study, we explore participants' perceptions of the resulting robotic coach and its interactions. We also particularly explored the design feature of \emph{affective adaptation}, which was considered as an important robotic coach feature in the previous design} study \cite{axelsson2021roman}. \hl{This study is part of a larger iterative, user-centred design process to create a robotic well-being coach} \cite{nielsen1993iterative}. \hl{As such, we analyse users' perceptions through qualitative analysis of post-study semi-structured interviews with the users} \cite{veling2021qualitative}, \hl{in order to contribute insights for the design of future robotic coaches}.

\subsection{Research Question}

\hl{In this paper, we aim to explore the under-studied area of robotic well-being coaching, by investigating the following exploratory research question through qualitative analysis:}

\begin{itemize}

    \item \textbf{RQ:} What are participants’ perceptions of the robot and interaction?

\end{itemize}

\hl{The exploratory research question aims to identify themes and patterns present in participants' perceptions of robotic coaching, in order to highlight new ideas, hypotheses, and areas of interaction with robotic coaches that should be further explored} \cite{swedberg2020exploratory}. \hl{To answer this exploratory question, we chose to conduct a qualitative analysis of the participants' perceptions, in order to generate qualitative insights for the future design of robotic coaches}, \hl{based on users' perceptions of our particular implementation}.  




\section{Related Work}
\label{sec:relatedwork}

\subsection{Well-being Coaching and Positive Psychology}

Coaching for well-being is a non-clinical approach that places an emphasis on the present, as well as how the coachee may thrive in the future and increase their fulfilment in life and at work~\cite{hart2001coaching}. Contrary to psychological therapy, which is intended to manage mental illnesses, coaching aims to boost the coachees' mental well-being, optimism, and goal-striving~\cite{green2006cognitive}. Coaching can emphasise various psychological practices, such as brief-solution focused practice (helping the coachee to focus on how their problem is already being addressed in small ways) and Cognitive Behavioural Therapy (focus on the relationship between thoughts, feelings, and actions), as well as Positive Psychology (paying more attention to the positive aspects of one's life). Different styles of coaching emphasise different psychological practices, such as cognitive behavioural coaching (examining the relationship between the coachees' actions, feelings, and thoughts), or brief-solution focused coaching (encouraging the coachee to notice and plan small steps that they are taking to address their problem)~\cite{green2006cognitive}.

\acf{PP} coaching is another style of coaching, which places emphasis on the positive aspects of the coachee's life, and fostering a positive attitude toward things that happen in their life)~\cite{seligman2007coaching}. Combining coaching with positive psychology exercises can improve the effect of positive psychology interventions~\cite{trom2022positive}. \ac{PP} practices can include focusing on positive experiences or positive individual traits~\cite{duckworth2005positive}. Other examples of positive psychology exercises can include writing down things the person is grateful for daily (in order to increase gratitude), and identifying a person's key strengths that they used a particular situation~\cite{meyers2013added}. Positive psychology has been widely used in organizations to enhance employee well-being and performance \cite{meyers2013added}, and in schools to improve student well-being, relationships, and academic performance~\cite{waters2011review}.

This work aims to apply pre-existing Positive Psychology exercises that have been found to be useful when conducted by a human coach, to robotic coaching where a Robotic Coach (RC) conducts these exercises with the coachees. The exercises used here are based on pre-defined exercises~\cite{cunha2019positive, emmons2002gratitude, gander2016positive, lopez2018positive}, and adapted to the robot together with a professional well-being coach. These exercises are described, in detail, in Section~\ref{sec:protocol}.

\subsection{Well-being in HRI}

Robots have recently been explored to address participant well-being. Positive psychology was previously explored in a study where students interacted with the Jibo robot over seven sessions at their home~\cite{jeong2020robotic, jeong2023deploying}. Participants showed improvements in their mood, well-being, and readiness to change. 
A study by~\citet{alimardani2020robot} examined the use of robot-led mindfulness meditation, finding that the participants' mood improved after the interaction. \textcolor{black}{\citet{matheus2022social} also examined mindfulness robots, conducting private deep-breathing sessions at a health centre with the custom-built robot Ommie. Finally, \citet{robinson2024brief} examined a robot instructing a deep breathing technique, with users reporting moderate usefulness and enjoyment in a randomised controlled trial. } Robots have also been used to improve well-being by instilling behavioural changes to help with weight loss~\cite{kidd2007robotic}, and to improve participants' mood by self-disclosing to a robot conversational partner~\cite{akiyoshi2021robot}. \citet{karim2022community} explored a community-based social robot that could help by displaying the users' mood data, and to raise community awareness about the emotions people feel. \textcolor{black}{Another study examined the use of a robot for conversation engagement, to support older adults' well-being, finding that cue-detection and emotion detection should be implemented to improve personalisation \cite{khoo2023spill}. \citet{rasouli2023co} identified design needs for an anxiety-reducing public speaking robotic coach, together with prospective users and well-being professionals. Similarly, \citet{alves2022robots} conducted a co-design study for a mental-health improving robot, together with the target group of adolescents.} 

However, most of these works follow static and scripted interactions or model limited adaptation without any personalisation towards individuals. To build on these previous works, our study examines the novel use of \emph{personalised affective adaptation in a robotic coach}. Adaptation and responsiveness, as well as the expression of empathy are both capabilites that participants would like to have in a robotic coach~\cite{axelsson2021roman}. \hl{As such, the implementation of the robotic coach in this work explores this design feature of affective adaptation.} 

\subsection{Affective Adaptation and Personalised Interactions}


Endowing social robots with affect perception mechanisms has been explored in several \ac{HRI} studies~\citep{McColl2015} aimed at improving individuals' experiences interacting with the robots. Robots are made to estimate user affective states by analysing their facial expressions~\citep{Li2020Deep}, body gestures~\citep{Kleinsmith2013,Noroozi2021Survey} and speech~\citep{Schuller2018SER} during interactions. This may enable robots to adapt their behaviours during interactions to appropriately reflect the users’ affective state, engaging them in meaningful conversations~\citep{Ficocelli2016Promoting, Churamani2022Affect}. However, this may not be easy in all situations owing to the unpredictability of human-centred environments resulting from changing environmental conditions and interaction contexts and the inherent uncertain and subjective nature of human affective behaviour~\citep{Dziergwa2018}. Most \ac{HRI} evaluations make use of \textit{off-the-shelf} affect perception mechanism that, despite providing state-of-the-art results on benchmark datasets, are not able to adapt to the dynamics of real-time interactions~\citep{Churamani2020CL4AR}. 
Enabling personalised affect perception, however, requires robots to adapt their perception towards individual users, accounting for individual differences in expression and other characteristic attributes~\citep{Churamani2020CLIFER}. Using such personalised models to encode the users' affective state as a contextual affordances, robots may dynamically model and adapt their interactions with the users instead of following the same static and scripted interaction for each user~\citep{Hemminghaus2017Adaptive, Churamani2017TheImpact}. Such personalised interactions are expected to enhance the users' experience interacting with the robot, offering a naturalistic interaction experience. 

Lifelong or \acf{CL}~\cite{THRUN199525, Parisi2018b}-based methods have been shown to be sensitive to dynamic shifts in data distributions and can balance novel learning in robots, as they interact with their environments, with the preservation and consolidation of knowledge across of past learning. For social robots, this entails learning during affective interactions with individual users, adapting their perception models to be sensitive to individual expressions~\cite{Gideon2017,pmlrbarros19a,Churamani2020CLIFER}. This allows for robots to offer \textit{personalised} interactions~\cite{Churamani2022CL4HRI} sensitive to individual behaviours, while retaining their generalisation abilities~\cite{Spaulding2021Lifelong} for a widespread application across different individuals. \hl{In the robot implementation in this work, we explore the use of both off-the-shelf non-personalised Adaptation (A), and Personalised Continual Learning -based (P) adaptation. For both, we use robots' affective appraisal of user's facial expressions during the interactions to modify the interaction flow by modelling additional dialogues based on the user's affective responses. For non-personalised adaptation (A), this is done using a pre-trained off-the-shelf affect perception model}~\cite{BCS2020FaceSN} \hl{while for the personalised adaptation (P) model, a CL-based model is used that personalises towards individual expressions}~\cite{Churamani2020CLIFER, Churamani2022CL4HRI}. \hl{Implementation details of the adaptation strategies can be found in}~\cite{Churamani2022CL4HRI}.

\subsection{The Proof-of-Concept Interaction Framework}
\label{sec:pocwork}

Our earlier \textit{proof-of-concept} work~\cite{Churamani2022CL4HRI} presented a multi-modal framework for enabling personalised \ac{HRI} capabilities in affective robots, especially in the context of \ac{PP}-oriented well-being coaching. \hl{In that prior study, we compared no adaptation to both off-the-shelf non-personalised affective Adaptation (A), and Personalised affective adaptation (P). We apply the same multi-modal framework from our prior study in the robotic coach implementation presented in this paper. The Personalised adaptation (P)} employs a \ac{CL}-based \textit{personalised} affect perception model~\cite{Churamani2020CLIFER}. The proposed framework allows for robots to be sensitive to individual affective behaviour during interactions and adapt the interaction based on such personalised perception. Our prior work \cite{Churamani2022CL4HRI} presented a \textit{proof-of-concept} user study comparing the proposed \ac{CL}-based \textit{personalised} interaction capabilities in the Pepper robot to using static interaction scripts or an off-the-shelf, deep \ac{ML}-based affect perception model~\cite{barros2020facechannel, BCS2020FaceSN}. The preliminary results indicated the participants' preference for \ac{CL}-based \textit{personalised} interaction capabilities in the robot, rating it high on \textit{anthropomorphism, animacy} and \textit{likeability} as well as on offering \textit{warm} and more \textit{comfortable} interactions, compared to static interactions, by being sensitive to the participants' expressions. The results presented in~\cite{Churamani2022CL4HRI} constituted only preliminary results. \hl{In this paper, we explore participants' impressions of the three different levels of affective adaptation (that is, None (N), affective Adaptation (A), and Continual Learning-based Personalised affective adaptation (P) from a qualitative perspective. Through this, we aim to further understand participants' perceptions of these different levels of affective adaptation.} 

\noindent 


\section{Materials and Methods}

\subsection{Robot and Study Design}

\label{sec:participatory_design}

We employed a user-centred and iterative design approach~\cite{nielsen1993iterative} when designing the robot's behaviour. To determine the design of our robotic coach and its interactions, we considered prior relevant works that studied what users would prefer in a robotic well-being coach, in particular our previous work,~\cite{axelsson2021roman}. \hl{In that work, we interviewed eight prospective robotic coach users and three professional coaches, to better understand what these stakeholders' expectations and needs for a robotic coach were. In this study, we designed the robotic coach interaction based on the findings of that design study, that is, we created an implementation iteration based on prior user-centred research. Our main focus was on exploring participants' experience of a Positive Psychology interaction with a robotic coach in a controlled laboratory environment. Additionally, we wanted to focus on exploring \emph{affective adaptation} in a robotic coaching interaction (as this is a feature that participants in the prior design study noted as important).} We opted to create a sense of \emph{empathy} in robot behaviour toward participants through affectively adaptive utterances, as suggested in ~\cite{leite2013influence}. \hl{Pepper's onboard speech is used to voice the utterances using the} \verb|NAOQi| \hl{\textit{ALTextToSpeech} framework.}
In order to create a salient script for the \ac{RC} where this empathy could be expressed through utterances, we collaborated with a professional psychologist (similar to~\cite{robinson2021robot}). The initial script was written by adopting \acs{PP} exercises appropriate for a one-off session~\cite{schueller2010preferences, o2012therapist}, with the assumption that the robot would have only a limited conversational understanding.
The script was improved (iterated upon) with the psychologist, by e.g. removing phrases that were deemed to be invalidating or patronising, improving the explanations for the exercises, clarifying prompts for participants to share their experiences, and determining the appropriate amount of conversation. This script was then implemented on the Pepper robot, which the psychologist interacted with in a $1$-hour session. During this session, the psychologist took notes on further improvements, which were implemented for the final version used in the laboratory study. 

\hl{In Adaptive (A) and Personalised (P) levels of affective adaptation, the robotic coach attempted to identify either a positive or negative valence in the user's facial expression during their response in the \textit{Two Impactful Things} exercise} (see Sec. \ref{sec:protocol}). \hl{If a positive expression was detected, the robot used a positive utterance, e.g. ``\textit{That sounds great, I'm happy for you.}'' or ``\textit{That's great, it sounds like a positive experience.}''. If a negative expression was detected, the robot used an empathizing utterance, e.g. ``\textit{That sounds like a tough experience. I'm sorry.}'' or ``\textit{I'm sorry that happened. That sounds difficult.}''. In the case of no affective adaptation (N), no such affective dialogues were used, and the robot uttered the neutral statement ``\textit{Thank you for sharing this with me.}''.}

\subsection{User Study}
\label{sec:studydesign}

\begin{figure*}[t]  
    \centering
    \includegraphics[width=\textwidth]{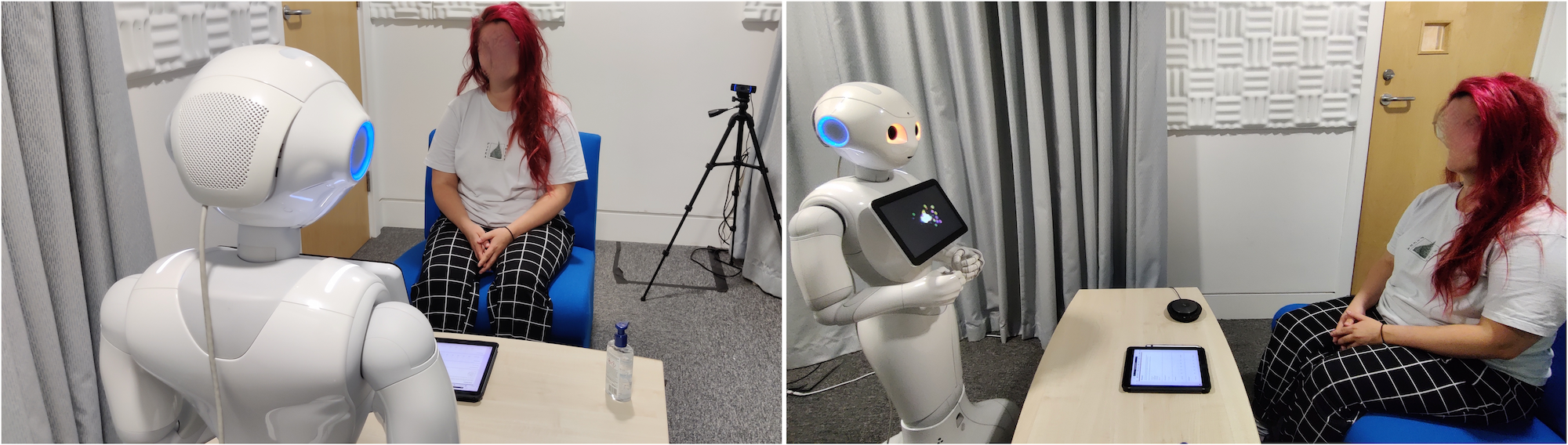}
    \caption{Participant and Pepper interacting with each other.}
    \label{fig:exp_setup} 
\end{figure*}

\subsubsection{Set-up}
\label{sec:setup}
As seen in Fig.~\ref{fig:exp_setup}, the participant sat in front of the \ac{RC} on a chair with a low table separating the two. A distance of $\approx1.0m$ was kept between the two to ensure that the \ac{RC} did not violate the participants' `personal space'~\cite{Mumm2011HRP, Mead2016}. Two cameras were placed facing the participant and the robot, respectively, to record the entire interaction. An external microphone was placed on the table in front of the participant to record their speech. Additionally, a tablet was placed on the table for the participants to fill out the  questionnaires. 

\subsubsection{Participants}
\label{sec:participants}
The user study was conducted as a \textit{within-subjects} study, \hl{that is, all participants experienced all three levels of affective adaptation. We chose this study design so that we could collect qualitative data from all participants about their perceptions of each of the adaptation levels.} We recruited $N=20$ participants ($12$ female, $5$ male, $3$ not disclosed) aged $26.70\pm3.68$ years from $12$ different nationalities, recruited amongst the students and members of the university. The majority of the participants ($N=13$) indicated little to no prior experience with humanoid robots. To specifically target a \textit{non-clinical} population, some of the enlisted participants were screened out if they were undertaking any mental health treatment or medication. All participants were asked to fill in the \ac{GAD7}~\cite{spitzer2006brief} and \ac{PHQ9}~\cite{kroenke2001phq} before they were enrolled in the experiments to make sure no participants were experiencing high anxiety or depression. Before the study, participants were also asked to watch two videos of the Pepper robot showing its interaction capabilities, in order to familiarise them with the robot. After the screening process, all participants provided \textit{informed consent} for their participation and the usage of their data for post-study analyses. The participants were compensated in the form of Amazon vouchers. The consent form, study-design and the experimental protocol was approved by the Departmental Ethics Committee.


\subsubsection{Experiment Protocol}
\label{sec:protocol}
The overall duration of the study was designed to be a maximum of $1$-hour,
depending on participant responses (average length was $50 \pm 11$ minutes, including the questionnaires after each interaction round). 
For each interaction round, the participants held similar interactions with the \ac{RC}, with only slight variations focusing on their \emph{past}, \emph{present}, and \emph{future}, in that order. This ordering reflects the emphasis that well-being interventions often place on imagining optimistic futures, which has been shown to reduce pessimism, negative affect, and emotional exhaustion~\cite{littman2014looking}. Expressly, for the past (first interaction round) the \ac{RC} asked the participants to recall events from the past few weeks, for the present (second interaction round); from the same day or week, and for future (third interaction round); they were asked to imagine situations that may arise in the coming weeks. The experiment protocol is described below:
%


\textbf{1. Introduction:} The \ac{RC} first greeted the participant by introducing itself, and explaining the experiment protocol. During the introduction, it asked yes/no questions making sure that the participant understood the instructions and were willing to interact with it. This allowed the participants to get comfortable with talking to the \ac{RC}. 
    
    \textbf{2. Administering the Experiment Conditions:} After the introduction, the \ac{RC} 
    assigned one of the three affective adaptation levels (None, Adaptation, or Personalised adaptation) to each round, that is, talking about the \textit{past}, \textit{present} and \textit{future}, respectively. \textcolor{black}{The condition orders were counterbalanced across participants, in order to avoid order effects. However, due to participants dropping out after conditions were already allocated, there was some imbalance across condition counterbalancing.} Depending upon the level of adaptation administered, the \ac{RC} aimed to sympathise with the participant, with empathetic utterances~\cite{leite2013influence} in accordance with the predicted participant affect.
    Each interaction round consisted of three exercises:
    \begin{itemize}
        \item [i.]\textit{Two Impactful Things:} In this exercise, the \ac{RC} asked the participants to talk about two impactful things or events that either happened in the past two weeks (\textit{past}), happened or are expected to happen today (\textit{present}) or are expected to happen in the coming two weeks (\textit{future}). The \ac{RC} also asked the participants to think about why these events happened and how do they made them feel. 
        \item [ii.] \textit{Two Things the Participant is Grateful For:} This exercise focused on developing \emph{gratitude}; a concept emphasised during \acsp{PP}, to increase positive affect, subjective happiness and life satisfaction~\cite{cunha2019positive,emmons2002gratitude}. The \ac{RC} asked the participant to recall or imagine (depending on \textit{past, present} or \textit{future} interactions) two things that they felt or might feel grateful for. 
        
        \item [iii.] \textit{Two Accomplishments:} In this exercise, the \ac{RC} asked the participant to think about \textit{past, present} or \textit{future} \emph{accomplishments}, focusing on self-esteem, which has been applied to increase well-being and ameliorate depressive symptoms~\cite{gander2016positive}. The \ac{RC} asked the participant to describe past, present or potential accomplishments, strengths the participant applied or may apply to accomplish these~\cite{lopez2018positive}, and how these accomplishments make the participant feel. 
    \end{itemize}

    During the study, despite following the same exercises, participants, on average, are witnessed to spend different amounts of time for past $(11 \pm 4$ minutes), present $(9 \pm 3$ minutes) and future $(10 \pm 3$ minutes) rounds with the overall coaching session taking $31 \pm 10$ minutes.

    \textbf{3. Interview and Debriefing:} After finishing all three interactions with the \ac{RC} a semi-structured interview was conducted with the participants, documenting their feedback and general impressions about the interactions. This enabled us to gather qualitative data on the participant's perceptions of the robot
    and debrief them on the objectives of the study and how the levels of adaptation were administered. The interview structure can be seen in Appendix~\ref{secA1}.



\section{Qualitative Analysis of Participants' Perceptions}
\label{sec:analysis_qualitative}

\hl{This section details the qualitative analysis procedure and results, focusing on examining and identifying themes about users' perceptions of the robotic coach and interactions (to address our research question), from a qualitative point of view.}
We applied Thematic Analysis (TA) \cite{clarke2015thematic} and analysed the data gathered from the post-study semi-structured interviews (see Appendix~\ref{secA1} for the interview questions). The 6 steps of Thematic Analysis consist of: 1) familiarization with the data (that is, transcribing and reading the data, and noting initial ideas), 2) creating initial codes, 3) identifying themes (that is, collating codes identified in step 2 into initial themes), 4) refining the themes (that is, reviewing the codes defined in step 3 in accordance with the identified themes and allocating them under each theme), 5) defining and naming the themes (that is, defining a description for each theme to describe the data), and 6) creating a report (for example, finding illustrative examples for each theme, and explaining them in a report). Participants were asked about their perceptions of the robot, interaction with it as a well-being coach, and its behaviour. We coded the transcribed interviews into ten major themes \hl{relating to our research question (``What are participants' perceptions of the robot and interaction?'').} The themes (see Fig.~\ref{fig:CL-HRI}) are presented below. \hl{After presenting each theme, we share insights for the future design of robotic coaches, based on the findings within that theme.}

\begin{figure}[t!]
    \centering
    \includegraphics[width=\textwidth]{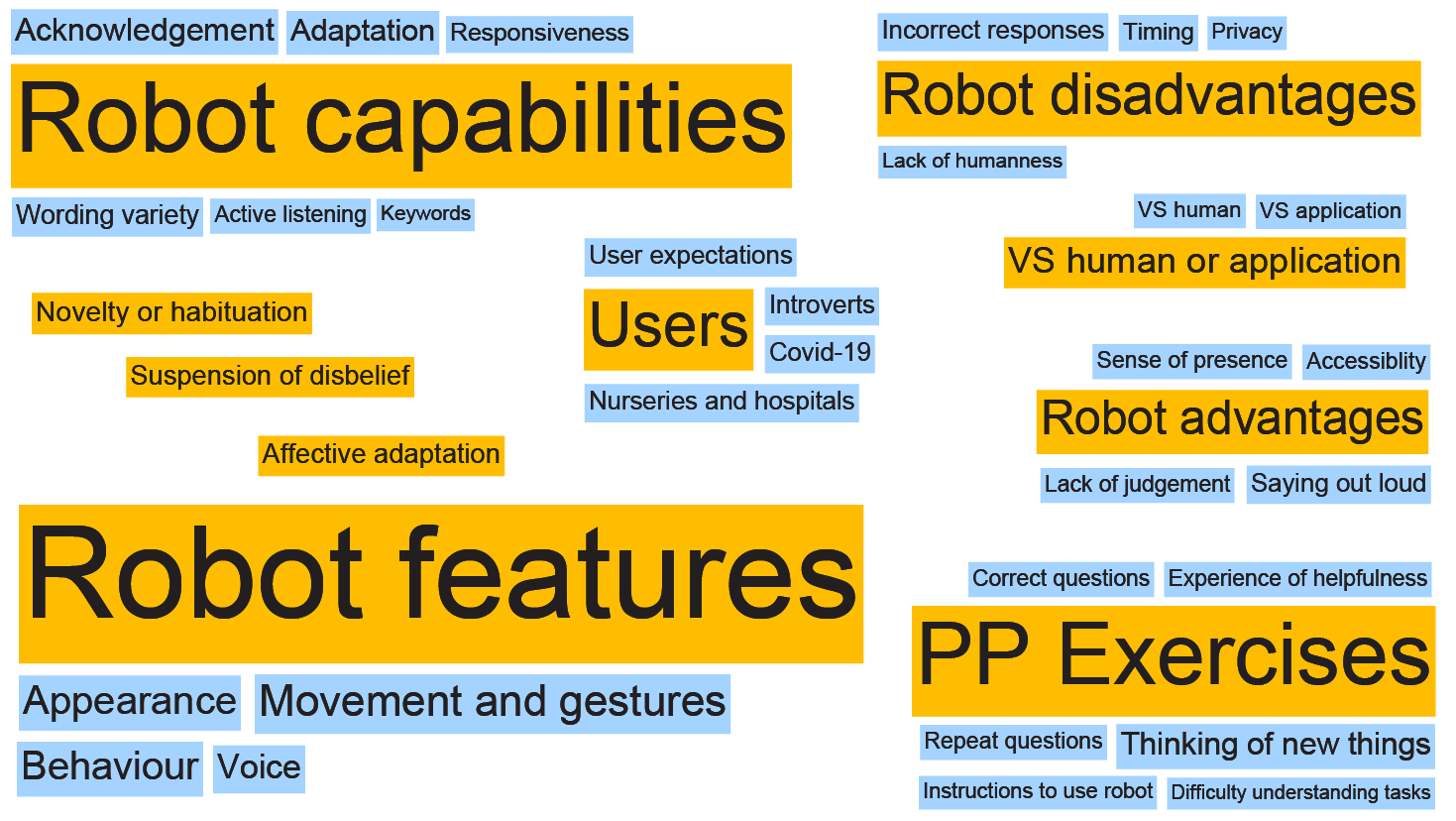}
    \caption{Themes defined in the TA are presented in orange, while codes related to these themes are presented in blue (best viewed in colour).}
    \label{fig:CL-HRI}
\end{figure}

\subsection{Positive Psychology Exercises}

Participants made several comments on the Positive Psychology (PP) exercises (see Table \ref{table:PPexercises}), noting that the exercises themselves were successful and helped them reflect on the positives in their life, and they experienced the robot being helpful. P9 underlined the importance of the robot asking the ``correct questions'', saying that working with a coach is important, and that the questions were well made in this case. P7 noted that they felt more grateful and positive after the exercise, and that the interaction was effective, especially those about the \emph{future}. Participants noted that the exercises made them \emph{think of new things}, but that some of them had \emph{difficulty understanding the tasks}. This could be mitigated with the robot further explaining the structure of the exercise session in the beginning. \emph{Instructions} on how to interact with the robot, as well as the ability to ask the robot to \emph{repeat questions} were called for in future interactions. Participants noted that improvements are needed on the robot's \emph{responses} (see Table \ref{table:PPexercises}). Participants also noted that the robot could be more \emph{responsiveness} in its responses, for example, P11 suggested that the robot could ask follow-up questions about telling someone about a thing they were grateful for, that is, who they would like to tell. This is discussed further in the next section (Section \ref{sec:capabilities}).

\noindent \textbf{Insights for Future Design of Robotic Coaches.} \hl{The robotic coach was seen by participants as successfully helping them reflect on the positives in their life, and recall new positive things. This indicates that Positive Psychology exercises are a promising avenue for future robotic coaching. In future works, more emphasis should be placed on the robot's capacity to clearly explain the tasks it is asking participants to do, e.g. by repeating the questions it is asking when a participant requests it. \textcolor{black}{Additionally, the capability for a robot to ask relevant follow-up questions would be helpful to users. Recent HRI research has investigated this, focusing on resolving ambiguities by enabling a robot to ask clarifying questions \cite{dougan2022asking}, and exploring morally sensitive clarification requests \cite{jackson2022enabling}.} }

\begin{table}[t]

\small
\centering
  \caption{Quotes from participants ($Pi, (i=1,2,...20)$) regarding the Positive Psychology exercises by the Pepper robot. $+/-$ signs indicate positive and negative statements, neutral statements do not contain signs.}
  {
  \begin{tabular}{|C|L|}\Xhline{1.0 pt}
    \rowcolor{gray!40}
    \textit{\textbf{PP exercises}} & \textit{\textbf{Quotes from participants}}\\ 
    \Xhline{1.0 pt}

    \rowcolor{gray!15}
    Experience of helpfulness   
    & 
    P2+: ``
    [It’s] an exercise we could do ourselves, but it felt more interactive because the robot was asking you questions.’’


    P6+: ``I particularly liked it trying to make me think about a  positive future. [...] That made me feel hopeful of what will come.’’
    


    P7+: ``At the end I think I felt more grateful. [...] I'm feeling more positive than I was before the start of the exercise, in that respect it was effective.’’
    
    P12+: ``If I [use it every day], it can potentially lift my mood. It's like a pet, but I'm not responsible for it when I travel or something like that.’’
    
    P19+: ``Pretty useful in terms of guiding me to reflect on certain things that I feel grateful for, it did have an impact on my present emotions, and I did feel better after the exercise.'' 
    \\
    \Xhline{0.5 pt}
    \rowcolor{gray!40}
    Correct questions   
    & 
    P2+: ``It was good that it explained the thought process behind the exercises [...] why thinking of the positive aspects was important.’’
    

    
    P9+: ``I think the questions were really well made, and so because of that it does help you centre a little bit, into ‘What has been positive?’, ‘What has been negative?’ It really felt a little bit like a coaching [or] counselling session. In that sense, it helped a lot.’’
    

    P3+: ``The questions really made me think, especially for the questions about future. The present and the past are usually something which we reflect on, but for the future it's something which is nice. It made me think about things that I usually don't.’’
    \\
    \Xhline{0.5 pt}
    \rowcolor{gray!15}
    Thinking of new things   
    & 
    P7+: ``It was good, trying to make me think ‘oh why would I feel grateful for things’ so that was that was good because I don't really think about that.’’

    
    P11+: ``It definitely made me think of stuff that I hadn’t thought of before, it took me a little while to think of things, and once I was thinking of them, I was like ‘oh yeah, let’s listen to that’ so I think it did its job.’’


    P4+: ``[...] It’s good to have the opportunity to think about these things. The difficulty sometimes is [being put] on the spot to list these things.’’

    P5: ``It feels weird to talk to a robot about anything personal at all. So I’m not sure I would choose that if I had an option. But surprisingly, at the end I was like ‘Oh yeah, I am grateful for things’, so I don't know. I guess it did work.’’
    

    P6+: ``I felt guided through things and [it] structured my thinking about different things.'' 
    \\

    \Xhline{0.5 pt}
    \rowcolor{gray!40}
    Repeat questions   
    & 
    P21: ``An ability to go back or repeat the question would fix a lot of the problems I had.’’
    
    P6: ` [...] I didn't understand one of the questions [...] so I asked if it could repeat that and I think it registered that as a next [answer]. So I guess [being able to ask it to repeat questions] would be better in making it feel like a conversation.''
    \\
    \Xhline{0.5 pt}
    \rowcolor{gray!15}
    Difficulty understanding tasks   
    & 
    P11: ``I couldn't quite see the structure ahead of time [...]. It asked what are you grateful for, and then when it talks about the accomplishments, I was like, 'oh, I totally should have used that'. [...] I just feel like it would have been more natural with a person to be like, 'oh, that one thing that I was just saying before'. And to a robot I kind of just felt like I said the same thing again as I did before, so I guess that was slightly weird.’’
    
    P16: ``These three [rounds (past, present, and future)], they all seemed similar. [...] It asks very similar questions. The first one was supposed to focus on the past and the second one on the present. And it asked similar questions like 'recent accomplishments' and that was confusing.''
    \\
    \Xhline{0.5 pt}
    \rowcolor{gray!40}
    Instructions to use robot   
    & 
    P2: ``Maybe I wasn't speaking loud enough, because I felt like I needed to repeat the commands on the feedback [portion of the interaction] a lot of times.''
    
    P3: ``I responded and it didn't get it, but after that it was very accurate. I didn't have to have any trouble, like having to wait for it. Sorry, you asked me to wait, but it didn't feel very natural.''
    \\
    \Xhline{1.0 pt}
  \end{tabular}
  }
\label{table:PPexercises}  
\end{table}


\subsection{Robot Capabilities}
\label{sec:capabilities}

\begin{table}[t]
\small
\centering

 \caption{Quotes from participants ($Pi, (i=1,2,...20)$) regarding the Robot Capabilities. $+/-$ signs indicate positive and negative statements, neutral statements do not contain signs.}
  \label{table:capabilities}
  {
  \begin{tabular}{|C|L|}\Xhline{1.0 pt}
    \rowcolor{gray!40}
    \textit{\textbf{Robot capabilities}} & \textit{\textbf{Quotes from participants}}\\ \Xhline{1.0 pt}

    \rowcolor{gray!15}
    Responsiveness   
    & {P5+: ``There were a few times when it did have a specific new response or it said ‘That sounds like a great accomplishment’ or ‘You should be proud’ and that was nice, I was like ‘Oh thank you robot for listening to me’.’’
    
    P18-: ``I would like to feel more of a difference between the reactions to something positive or negative I’m saying. If I said ‘I'm very happy’ I wouldn't like to say ‘That's fine’. [I would like it to] engage more positively with my positive feelings.’’
    
    
    
    
    P10+: ``Even though it was pretty much always the same response, I think it's nice that she responded to everything I said.''
    



    
    
    
    P1-: ``It wasn't really useful past the fact that it did ask the questions [...]. If there’s a complex robot in front of me, in order to do more than just readouts, it would be nice if it could respond.’’
    }
    \\
    \Xhline{0.5 pt}
    \rowcolor{gray!40}
    Acknowledgement
    & {P14-: ``You say something and then it continues along not really acknowledging what you said.’’

    P20-: ``[...]  [the thing that made] me feel that it wasn't useful was when I didn't feel that what I was saying was being reacted to in a way I expected.’’

    P13: ``It did ask good questions, but then there wasn't any room for follow up.’’


    
    P1: ``If you did add acknowledgement to it while [someone is] talking, [...] it would reinforce the person’s belief in the robot actually understanding what you're saying. As opposed to when it's just standing, then I don't know if it actually really understands what I'm saying.’’
    }
    \\
    
    \Xhline{0.5 pt}
    \rowcolor{gray!15}
    Active listening
    &
    P13: ``It asked good questions that made me think. But it felt more like the robot has a good script than the robot is engaging in active listening.’’


    P19-: ``Sometimes I feel like [it's] not really listening because [there are] very few responses, [and it's] choosing between the available sentences. So [it] can say ‘Umm, I see. Thank you for sharing that with me.’, with exactly the same [tone] every single time.'' 

    P10: ``More natural head movements and arm movements, while I'm talking. [...] If that's enhanced, [it's] useful for keeping me engaged. [...] Nodding or saying ‘Umm’, ‘OK’, or ‘Wow’.’’

    P8+: ``I think the head nods were quite real and if it gets very in sync, those were one of the things that made me feel like I was being listened to.’’
    \\
    
    \Xhline{0.5 pt}
    \rowcolor{gray!40}
    Wording variety
    & 

    P16: ``Maybe include more of a variety of responses, [so it’s] not always the same.’’

    P11: ``I guess a wider repertoire of different things it might do [...]. There were certain phrases it would say lots of times. I think it was moving the same way every time.'' 

    P6: ``[To feel that it] understood how I felt, it would have required more of a range of replies.’’

    P4: ``If there is a bit more variability, it would feel more natural. [...] Because then when you do this five times [...] it loses all sense of humanity. You just disconnect because you can see the automaton. Now you can see the machine there.’’
    \\
    
    \Xhline{0.5 pt}
    \rowcolor{gray!15}
    Adaptation
    & 
    P13-: 
    ``[...] [it was] asking me questions that are making me think productively on my end, but it’s not really responding to what I’m saying.’’


    P9: ``[...] I don't know how much it actually was just reacting to keywords, or was really sensing what I was feeling.’’
    
    
    P3-: ``It felt a bit inert [...] it can give some suggestions, like ‘try focusing more’.'' 
    \\

    \Xhline{0.5 pt}
    \rowcolor{gray!40}
    Keywords
    &
    P12: ``I told the story, so maybe mentioning some keywords from my stories rather than using a template [...].’’
    
    P17: ``I thought it could do a little bit more personalization, if I say the the keywords, like 'organisation' or 'events' it could say 'Oh, I hope you have a nice event' or [...] 'You have a nice organisational skills' or 'Seems like good organisational skills', so it can pick up words and make it seem more human interactive I think.''
    \\
    
    \Xhline{1.0 pt}
    
  \end{tabular}
  }
\end{table}

Robot capabilities were also discussed outside of the PP exercise content itself (see Table \ref{table:capabilities}). As noted in the context of PP exercises, participants wanted more \emph{responsiveness}, noting that the robot was ``inert'' in its lack of suggestions (P3). P5 noted that they enjoyed the robot giving specific new responses such as ``that sounds like a great accomplishment'' and ``you should be proud'', describing a sense of being listened to. This \emph{wording variety} should be emphasised in future applications of a robotic coach. A sense of \emph{acknowledgement} and \emph{active listening} (a technique used in coaching \cite{bresser2010coaching}) could be emphasised further by picking up \emph{keywords} from the participants' speech, and \emph{adapting} the robot's responses specifically to these. P1 noted that this would ``reinforce the person's belief that the robot was understanding what they're saying''. 

\noindent \textbf{Insights for Future Design of Robotic Coaches.} \hl{A key area for future robotic coach design is creating more responsive robot utterances and follow-up questions, in terms of their content. This could help participants feel more acknowledged and actively listened to, which is considered important in coaching }\cite{bresser2010coaching}. \textcolor{black}{Active listening  techniques include backchannelling (e.g., ``mm-hmm'' vocalisations),  and paraphrasing the speaker. Future research should examine how such techniques could be used with a robotic coach.} \textcolor{black}{Other active listening techniques include relevant follow-up questions and paraphrasing the speaker \cite{fitzgerald2010active}.} \hl{Future research should explore e.g. applying generative methods such as Large Language Models (LLMs) to generate follow-up questions that are adapted to the content of users' utterances, a research avenue which has been extensively explored in HRI after the completion of the user study presented in this work} \cite{zhang2023large}. \textcolor{black}{Accordingly, this finding on the importance of responsive utterances has been underpinning our subsequent works that investigated the use of LLMs in robotic well-being coaching. For instance, in our follow-up work, we have investigated how LLM-generated language is perceived in robotic coaching \cite{spitale2024appropriateness}, and how LLMs can be embedded in an autonomous robot system \cite{spitale2023vita}.}


\begin{table}[t]
\small
\centering
  \caption{Quotes from participants ($Pi, (i=1,2,...20)$) regarding the robot's features. $+/-$ signs indicate positive and negative statements, neutral statements do not contain signs.}
  \label{table:features}
  {
  \begin{tabular}{|C|L|}\Xhline{1.0 pt}   

    \rowcolor{gray!40}
    \textit{\textbf{Robot features}} & \textit{\textbf{Quotes from participants}}\\ 
    \Xhline{1.0 pt}
    \rowcolor{gray!15}
    Behaviour
    & P2+: ``Very supportive, especially once it starts to properly respond to the content of what we’re saying. [...] very helpful, it kind of guided you through the whole process.'' 
    
    P2-: ``I think at some points it felt a bit condescending. Maybe it's just my ego talking, it just felt a bit weird to be told ‘well done’ by a robot.'' 

    P5-: ``It accomplished the task, but it was robotic [...].’’

    P10+: ``A lot of listening and giving space to talk. I think that’s important.’’

    P9+: ``It made me feel comfortable, so I was OK in talking about these things.’’


    P8+: ``One really nice thing was that I felt heard. 
    [...] and I felt like we were interacting.’’


    P16+: ``It gave me a good feeling because it seemed compassionate.’’

    P16-: ``I felt a bit stressed out because I knew that it would respond quickly and I'm not very fast when it comes to classifying emotions.’’



    P20+: ``I liked that it stays calm. When someone talks to another person who isn't a qualified professional, it's very hard to get someone to stay completely calm in the face of some things.’’



    P4: ``I like the friendliness, it’s a good thing. I like the positive reassurance that it gives when it talked to me about the achievements and then said 'great you deserve this', that's good to hear.'' 
    \\
    \Xhline{0.5 pt}
    \rowcolor{gray!40}
    Voice
    & P19+: ``[The robot] speaks at a relatively slow pace,[...] it’s patient and listening.'' 

    P9+: ``The tone of voice was also really good.’’

    P14-: ``I was very creeped out. The technical speech engine isn't human.'' 

    P20+: ``The voice is pleasant. There's some weird pronunciation [...], but otherwise it’s good.’’

    P21+: ``I like the voice actually. I'm cool with this slightly robot-like inflection.'' 
    
    P7: ``I like the friendly like tone voice. That made me want to talk more with the robot.’’


    P3: ``It has good, pleasant voice.’’
    
    P6: I like the voice. It doesn't seem as mechanical as I thought it would.’’
    \\
        \Xhline{0.5 pt}
    \rowcolor{gray!15}
    Appearance
    & 
    P9+: ``There's a lot of things here that together all of it may be a very very positive figure. And it doesn't feel very threatening at the moment. [...] It’s very cute.'' 

    P21: ``It’s just about is at the high end [of being humanoid] before it gets too creepy. [...] It didn’t have to be an entire robot, [it could be] a rectangle with eyes maybe.’’


    P10+: ``The human form factor is quite nice.'' 

    P21+: ``It does have quite a friendly face, it's non-threatening. It’s a good size.'' 

    P7: ``I think if it was more human-like, that would actually have a negative impact.’’


    P15+: ``That it didn't have any expressions [...] was actually good. It was like I'm talking to somebody, but at the same time that person is [...] just taking in what I'm saying, it's not really even giving back. That's actually a good thing.’’
    
    P20-: ``Because the expression is the same, it's very hard to see if the robot is empathising.’’
    

    P4+: ``I like that it has a more positive than neutral face. It's more approachable like that.''
    \\
    \Xhline{0.5 pt}
    \rowcolor{gray!40}
    Movement
    & 
    
    P17: ``During speaking it can make some motions. [...] There are other ways of agreeing and not agreeing in body language terms. Tapping, clicking, thinking, putting their hands together in a more lifelike in a way.’’


    P1-: ``There's nothing that's clearly missing, except in terms of movement. Maybe moving around while I'm talking as opposed to only when it's responding. And acknowledging things as I talk.'' 

    
    
    P14+: ``The way it moves is actually not too odd. The movement seems to be fairly lifelike.’’
    
    P21-: ``Because it was always mostly the same gesture, that gesture made it artificial.’’

    P19-: ``[It's] trying to show open, welcome, and friendliness, but we don't really get the meaning. It's always the same movement with exactly the same angle, so it's easy to see the programming.'' 
    

    P18: ``Even though I like the expression of the hands, it could be a little bit repetitive.’’
    \\
    
    \Xhline{1.0 pt}
  \end{tabular}
  }
\end{table}

\subsection{Robot Features}

Participants made notes on the robot's features (Table \ref{table:features}), such as its \emph{form}, \emph{voice}, \emph{movements} and \emph{behaviour}. In appearance, P9 described the robot as an overall ``positive figure'' and ``cute''. P21 noted that it was at the high end of humanoid before being creepy, while P15 called it ``non-threatening'', P4 ``approachable'', and P10 ``animated'' as it spoke. Overall, the Pepper robot was evaluated positively on appearance for this task. Participants made a few specific complaints about Pepper's appearance, noting that they were sometimes distracted by the light in its eye, and were sometimes confused by the function of Pepper's tablet displaying a screen saver animation. However, some participants noted that they enjoyed the animation and it made them feel relaxed. For future interactions, a robot with no eye light or tablet could be considered, if they are not relevant for the interaction.

Participants had more feedback on its \emph{movement}, calling for more gestures to signal \emph{body language} and \emph{turn-taking}, making it clearer when they should be speaking. Gestures were also seen as ``artificial'' and ``repetitive''. This could be mitigated by developing a more sophisticated method for back-channeling during conversations.


\begin{table}[t]
\small
\centering
  \caption{Quotes from participants ($Pi, (i=1,2,...20)$) regarding the advantages of a robotic coach. $+/-$ signs indicate positive and negative statements, neutral statements do not contain signs.}
  \label{table:robot_adv}
  {
  \begin{tabular}{|C|L|}\Xhline{1.0 pt}   
    \Xhline{0.5 pt}
    \rowcolor{gray!40}
    \textit{\textbf{Robot advantages}} & \textit{\textbf{Quotes from participants}}\\ 
    \Xhline{1.0 pt}

    \rowcolor{gray!15}
    Saying out loud
    & 
    {P21+: ``Being forced to do it out loud is helpful.’’

    P14+: ``It was good to kind of think about what's going on at the moment’’

    P15+: ``[...] When you think about it and you say it out loud, it puts a lot of things in perspective, so I think the robot did help me do that.’’

    P4+: ``[...] you can still benefit from simply having to say things, and listening to them as well.’’
    }
    \\
    \Xhline{0.5 pt}
    \rowcolor{gray!40}
    Lack of judgement
    & {P3+: ``It’s easier to talk to a robot somehow than a real human being. You don’t face judgement from a robot, so that’s nice. It’s an open ended question so sometimes you feel stupid saying certain answers to a human than a robot.’’

    P19+: ``It does a good job in listening and not judging you.’’

    P15+: ``I felt like I wasn't getting judged. [...] the fact that it didn't have any expressions was actually a good thing. A lot of times when you talk about things you look at another [person], whether verbally or non-verbally [they react] [...], and the fact that it didn't do it made me feel like I can say as much as I can.’’
    
    P8+: ``When you're trying to do this with a person one more wall that you have to navigate, which I didn't feel like in the beginning. [...] You're not entering a room and sitting with a person, so there’s less social layers you have to navigate.’’
    }
   \\
   \Xhline{0.5 pt}
    \rowcolor{gray!15}
    Accessibility
    & {P9: ``For coaching it could definitely help. Not for making a first assessment, because I think that still should be a human. But needing someone that you can interact with and actually go through the things you’re [thinking about], having the robot react to this, […] that's really good.''
    }
   \\
   \Xhline{0.5 pt}
    \rowcolor{gray!40}
    Sense of presence
    & {P1+: ``It's obviously not going to be on the same level as as a human [...] so the question really is, is it better than just something you can do to yourself? [...] If it were online then [...] you can just go to any question and just skim through it. Whereas with this, you actually have to sit down and it has to actually say each one to you. So you're forced to consider each question and in that sense I suppose it's better than just reading something.’’
    
    P20: ``It's like seeing an online quiz, [...], but having a text to speech sort of thing and tell me, that and having a physical presence in the room with me. That establishes more of a connection. It's not the same as if it were a human.''
    }
   \\
    
    \Xhline{1.0 pt}
  \end{tabular}
  }
\end{table}

In its behaviour, the robotic coach was called ``very supportive'' (P2), ``compassionate'' (P16), and ``giving space to talk'' (P10), all important qualities for a robotic coach. However P2 noted that it also felt ``condescending'' at points, even if the robot telling them ``well done'' was intended to be friendly. The robot's encouragement phrases should be further evaluated on when they are useful, and when they are seen as too invasive. This could vary according to participant, which makes examining how different personality traits affect perceptions of the robot important. 

Participants also commented on the robot's \emph{voice}, appreciating the ``slow pace'' that made it appear ``patient'' (P19), its \emph{tone of voice} (P9), ``pleasant'' (P20), and ``friendly'' (P7). P14 however said they were very ``creeped out'' by the voice, and P13 called it ``weird''. This illustrates that there is no one-size-fits-all solution with regards to the voice of a robotic coach --- rather that the voice should be selected in order to be congruous with the expectations of a wellbeing coach (patient, pleasant, and friendly).

\noindent \textbf{Insights for Future Design of Robotic Coaches.} \hl{The Pepper robot's appearance and voice were generally perceived positively by the participants. This indicates that Pepper and its onboard voice are suitable for robotic coaching. However, future research should explore whether participants' ability to select the voice (customization) might positively impact the coaching experience, as some of them did not enjoy it. Prior work in voice agent interactions found that customization via personality-matching increased their attractiveness} \cite{snyder2023busting}. \hl{The robot's movement received negative feedback (e.g. ``artificial'' and ``repetitive''), showing a need for future research into more naturalistic robot gestures that support turn-taking during conversation. In prior work, gestures were described as one of the methods with which turn-taking in HRI can be improved }\cite{skantze2021turn}. \hl{The robot's verbal behavior was also perceived generally positively (e.g. ``compassionate'' and ``supportive''), but at some points ``condescending''. Future research into users' perceptions of the design of specific phrasing of encouraging phrases should be conducted to examine how phrasing could be improved. } \textcolor{black}{Influenced by this finding, our subsequent work have been investigating the appropriateness of robotic coach language in well-being coaching, examining the themes of efficiency and time pressure, empathy and emotions, as well as bias and ethics \cite{spitale2024appropriateness}.}

\subsection{Robot Advantages, Disadvantages and Comparisons with Human or App-based Coaching}

Participants detailed both robot advantages (see Table \ref{table:robot_adv}) and disadvantages (see Table \ref{table:robot_dis}). Participants also made favourable and negative comparisons of the robot to both a human coach and a hypothetical phone or computer application performing the same PP exercises (Table \ref{table:comparison}). These three themes are interrelated, and are all discussed in this section.

\begin{table}[t]
\small
\centering
  \caption{Quotes from participants ($Pi, (i=1,2,...20)$) regarding the disadvantages of a robotic coach. $+/-$ signs indicate positive and negative statements, neutral statements do not contain signs.}
  \label{table:robot_dis}
  {
  \begin{tabular}{|C|L|}\Xhline{1.0 pt}   
    \rowcolor{gray!40}
    \textit{\textbf{Robot disadvantages}} & \textit{\textbf{Quotes from participants}}\\ 
    \Xhline{1.0 pt}

    \rowcolor{gray!15}
    Timing
    & P19-: ``It can be a distraction, when I haven't finished the whole speech, but I'm just pausing a little bit, and [it] thought I finished and then [it] asked me questions.'' 

    P10: ``The long pauses [were awkward], but you also don't want to interrupt people.’’

    P10-: ``While I narrate saying ‘Oh wait, I’m thinking’, [the robot] thinks I've said something and goes on to the next question. [...] I didn't know what to do to give myself time to think.’’

    P9-: ``I didn't understand the interval it was giving me to talk, if it’s a fixed interval. Sometimes it would interrupt me or I almost couldn't finish the sentence and it was asking the next question.'' 

    P20-: ``There were times that I felt interrupted, so there was a point when I was thinking about something. [The robot says], ‘oh I couldn't understand what you said’, so it's that sort of speech processing and recognition. So at times it felt that it was trying to hurry me up.'' 
    
    P18: ``Sometimes you want to reply saying 'thank you' as well. And I don't know if the robot is waiting for me to say that or not.’’
    \\
    \Xhline{0.5 pt}
    \rowcolor{gray!40}
    Incorrect responses
    &
    P6-: ``I was saying I managed to travel and it was the first time that I’ve done that in a while, and just spent time with people I care about. And the replies were. ‘That sounds like a tough experience’, so I don't know what triggered it to say that, but it was the complete opposite of what I was feeling and it was making me feel obviously misunderstood. And also made me question what was I saying that was making this happen.’’
    
    P15-: ``I think at times I would say something positive and it would say ‘oh, I'm sorry you felt that way’. […] At times, it didn't understand what I was trying to say.’’
    \\
    \Xhline{0.5 pt}
    \rowcolor{gray!15}
    Privacy issues
    &
    P4: ``It's hard to choose what you even say, especially if you know you're being recorded, you have to choose something that is not going to expose you extremely at the same time.''
    
    P19: `` I feel like more like I'm being recorded, than being heard, or [listened to].''
    \\
    \Xhline{0.5 pt}
    \rowcolor{gray!40}
    Lack of humanness
    &
     P7-: ``If this was an actual person asking me the same questions, maybe it would have had a better impact like because maybe I could have had a more personal connection with them. And the robot is quite limited [in] the depth of information that it can can ask. You can't program it to ask follow up questions. A real therapist or  [...] a coach [...] could have done better [...].’’
     
    P19: ``It doesn't have some of the bad things a human coach would have, like negative comments. [...] On the flip side [the robot] didn't give me an organic response. It’s always the same response and sometimes it can be demotivating.’’
    \\
    
    \Xhline{1.0 pt}
  \end{tabular}
  }

\end{table}

\begin{table}[t]
\small
\centering
  \caption{Quotes from participants ($Pi, (i=1,2,...20)$) comparing the RC to a human or another type of application. $+/-$ signs indicate positive and negative statements, neutral statements do not contain signs.}
  \label{table:comparison}
  {
  \begin{tabular}{|C|L|}\Xhline{1.0 pt}   
  \rowcolor{gray!40}
    \textit{\textbf{RC VS human or application}} & \textit{\textbf{Quotes from participants}}\\ \Xhline{1.0 pt}
  
    \rowcolor{gray!15}
    RC VS human
    & 


    P10: ``Not too bad, [it’s] obviously not a replacement for [a human]. If there's no human counsellors or coaches available, then it's possible that it can have a good effect on people.’’

    P9+: ``I wonder how different this would be from a real human session. It made me feel comfortable, so I was OK in talking about these things. So in that sense, it's definitely positive.’’




    P5-: ``It was appropriate in telling me what to think, but it didn't seem like it was understanding things because it just said the same thing back. Which is fine, because it's not actually my therapist, but probably would be irritating if you were really pouring your heart out.’’
    \\
    
    \Xhline{0.5 pt}
    \rowcolor{gray!40}
    RC VS application
    &
    P20+: ``I've tried this in the past as well, maintaining a gratitude diary. [...] When someone puts you on the spot to think about it's more effective, than trying to enforce it myself.’’

    P21+: ``The voice, the interactive part of it, that is different to doing it on a screen, or in self help books. So I think that is positive.’’
    
    P10+: ``It’s quite useful, and a better format than doing it on a laptop or something like that.’’

    P9+: ``We've seen some of these coaching exercises through a computer when you even have just a sentence, like let me think about the positive thing. So having a voice and a sort of reaction is a really good step, so I think in that sense as a coach, I think it was cool. It was very positive.’’
    \\
    
    \Xhline{1.0 pt}
  \end{tabular}
  }

\end{table}

A major advantage that participants saw in using the robot was \emph{speaking out loud}. Participants felt that verbalizing their thoughts helped them \emph{think of new things} and put things into perspective, and benefited simply by hearing themselves talk. This is supported by literature, where self-talk is found to be related to motivation \cite{hardy2006speaking}, and that the \emph{production effect} indicates that words spoken out loud improve explicit memory in comparison to words simply read quietly \cite{macleod2010production, icht2014production}. However, speaking out loud can come with the disadvantage of \emph{incorrect robot responses}. On occasion, the robot would respond incongruously to an emotional experience that a participant had described, making them feel misunderstood. This suggests that there is a delicate balance in using adaptive robot utterances, so that participants can benefit from speaking and being responded to, while mitigating the negative effects of incorrect responses. The same negative effects can be seen in the disadvantage of \emph{incorrect timing}: some participants noted that long pauses were awkward, or that the robot would begin responding when the participant hadn't finished. This could be improved by enhancing \emph{turn-taking} ques, e.g. with robot backchannelling via vocalizations and gestures.

Another such delicate balance in robotic coach design is the relationship between \emph{privacy and data collection}, as well as \emph{adaptation} and \emph{responsiveness}. Multiple participants wished the robot would adapt to their speech in its verbal utterances, creating \emph{wording variety} with picking up \emph{keywords} from their speech. This could improve the impression of \emph{active listening}, which they saw could be a potential advantage over a mobile phone or computer application. However, participants were concerned about being recorded while sharing their emotional experiences. The concern for lack of privacy could be mitigated through \emph{user education} on what data exactly is being collected, and how it is being stored.

Participants also commented on both the robot's \emph{lack of judgement}, but also its \emph{lack of humanness}. Participants noted feeling comfortable with the robot, and that there were less social layers to navigate than with a human coach. A participant also noted that it didn't have some of the ``bad things'' that a human coach may have (such as negative comments), however that the responses weren't organic, which could be demotivating. On the other hand, some participants experienced a \emph{lack of judgement} from the robot, noting that it was easier to talk to, and that they ``didn't feel stupid'' giving the robot answers, as they might feel with a human coach. A participant noted that the robot's lack of facial expressions was a good thing, and that they felt like they could ``say as much as they can'' without reactions from the robot. This suggests that robotic coaches are useful for certain use cases, but could be irritating if they were ``pouring their heart out''.

\noindent \textbf{Insights for Future Design of Robotic Coaches.} \hl{Participants perceived that speaking out loud with the robotic coach was one of its key advantages. \textcolor{black}{This is in line with research on the ``production effect'', that is, individuals better recalling information that is spoken out loud, rather than read \cite{macleod2010production}. Even speaking out loud to one's self (``self-talk'') has been found to be useful for motivation \cite{hardy2006speaking}.} The other side of the coin with speaking out loud was that the robot made conversational errors such as interrupting the user, which have been described as common social errors in prior work examining a taxonomy of robot errors} \cite{tian2021taxonomy}. \hl{Future research should investigate how to mitigate these errors, and how to design repair strategies for a robotic coach to recover from such errors.} \textcolor{black}{As a first step towards implementing this recommendation, we have started to investigate users' behavioral responses to robot errors \cite{spitale2023longitudinal}, as well as the use of robotic repair strategies for social errors made by a robotic coach in a longitudinal interaction \cite{axelsson2024oh}.}

\hl{Another key area of investigation is privacy in robotic coaching, which was one of the main concerns of users in terms of disadvantages of a robot. Future research should investigate user perceptions of which information is appropriate for a robotic coach to sense, store, and adapt to, as well as investigate how to appropriately educate users on their privacy when interacting with a robot.} \hl{Privacy in HRI has been explored in prior work} \cite{rueben2016privacy}, \hl{but cohesive privacy guidelines for users do not yet exist. Such guidelines should be developed in future research to inform both robot developers and robot users on best privacy practices in HRI.} \textcolor{black}{This finding has motivated us to focus on developing guidelines for addressing ethical considerations, including privacy with users of robotic coaches \cite{axelsson2024robots}}. 

\hl{Participants perceived the robotic coach positively for its ``lack of judgement'', indicating that robotic coaches can be a promising avenue for participants to express themselves. However, participants also noted that sharing a particularly personal experience with a robotic coach could be frustrating due to its ``lack of humanness'' (lack of emotional expression and lack of responsiveness).} \textcolor{black}{Future research should investigate how to educate users of robotic coaches about their limitations. Our ongoing research work brings this recommendation into life by examining how to educate users on what robotic coaches are ``good at'' (that is, facilitating non-clinical well-being exercises) and what they are ``bad at'' (that is, conducting meaningful therapeutic conversations) \cite{axelsson2024robots}}.


\subsection{Users}
\label{sec:user_expectations}

\begin{table}[t]
\small
\centering
  \caption{Quotes from participants ($Pi, (i=1,2,...20)$) regarding the potential users of the robotic coach. $+/-$ signs indicate positive and negative statements, neutral statements do not contain signs.}
  \label{table:users}
  {
  \begin{tabular}{|C|L|}\Xhline{1.0 pt}   
  \rowcolor{gray!40}
    \textit{\textbf{Users}} & \textit{\textbf{Quotes from participants}}\\ \Xhline{1.0 pt}
  
    \rowcolor{gray!15}
    User expectations
    &
    P6: ``I guess it's more beneficial for people who have never done therapy or that kind of stuff. Because if you're already used to laying things up like this and focusing on the positive aspects rather than the negative ones, then it feels very obvious, what it's doing.’’


    P3: ``I think it also depends upon the expectations of the people. So if some days I'm feeling really crappy and if I'm expecting more, maybe then I wouldn't get what I want.’’

    P3: ``It works to at the very basic level. It definitely works, but you don't have the same day every day. So it depends upon that. What kind of expectations you have coming into this.’’
\\
    \Xhline{0.5 pt}
    \rowcolor{gray!40}
    COVID-19
    & 

    P12: ``[With] COVID quarantine [...] if this is industrialised for people who are isolated at homes [...]. It's a moving intelligence, and being able to understand your difficulties, and trying to help with some mental health problems. I think that's really, really beneficial.’’

    P12: ``If we apply this to the home, to a domestic situation. I would say it’s very useful. [...] even in not Covid-19 times, I think it's still very useful. And for these people who are at care homes or or these people who are not sociable, like in Japan, a lot of teenagers are absorbed in computer games and they only talk to virtual machines. So I would say this is a very good channel for them to open themselves back to society, like a transition period Because the computer does interact. So this can bridge them from self isolate situation to a more [social life].’’
    \\
    \Xhline{0.5 pt}
    \rowcolor{gray!15}
    Introverts
    & 
    P12: “For an introverted person it's really helpful [...] [you] could be acting very shy to a real person, but I feel quite relaxed to talk, because in your mind it's still a machine.”

    P9: ``It will be helpful for a lot of people who might have awkward interactions with humans.’’

    P9: ``[...] People who feel awkward when they want to say things that are very personal and private. If the robot can react, and help people go through their problems, it’s a huge thing.’’
    
    P17: ``Some people feel more confident talking on video call [...] this might cater to a particular part of the of people who who are more confident speaking to a robot.'' 
    \\
    \Xhline{0.5 pt}
    \rowcolor{gray!40}
    Nurseries and hospitals
    &
    P9: ``[...] that could help a lot in nurseries and hospitals. You can't have a nurse to every person, but you could have a robot. Having these small interactions, this could make a huge difference.’’

    P9: ``[...] my parents worked in the health sector and I know these can provide a huge help. [...] especially thinking about older generations who are alone [...].’’
    \\
    \Xhline{1.0 pt}
  \end{tabular}
  }
\end{table}


Participants also made suggestions as to for whom such a robotic coach could be useful (Table \ref{table:users}). Participants saw that the robot worked on a basic level, and could work especially for users who didn't have a lot of experience with PP exercises, and could benefit from learning them and the positive thinking from the robot. However, high expectations might lead to disappointment. Participants suggested that \emph{introverted} users may benefit from the robot, if users were to feel shy or awkward talking to a human. A participant specifically mentioned the robot being useful at their home during Covid-19 social distancing conditions. Another user suggested the robot could be useful in the \emph{health sector}, such as in nurseries and hospitals, where there is a lack of staff, and having small interactions with a robot could make a positive difference. One participant mentioned that it ``depends on the expectations of people'', suggesting that deploying these robotic coaches into the field needs to be carefully framed to participants as to their usefulness and capabilities. 

\noindent \textbf{Insights for Future Design of Robotic Coaches.} \hl{Participants mentioned that robotic well-being coaches could be helpful in particular contexts or circumstances where people feel isolated, such as healthcare (where patients may experience a lack of company to speak with), or during a a pandemic (e.g., COVID-19 where social distancing measures were followed). \textcolor{black}{These findings align with previous research in HRI.} For instance, Ghafurian et al. found that people's perceptions toward social robotic companions changed to be more positive after the Covid-19 pandemic, and that feelings of loneliness or isolation could drive the purchase of a social robot} \cite{ghafurian2021social}. \textcolor{black}{Similarly, \citet{laban2024building} found that participants reported decreased feelings of loneliness after a conversation with a social robot, during the COVID-19 pandemic.} Future research should further investigate the capability of robotic coaches to contribute positively in such circumstances. 

\hl{Participants also mentioned that introverted people in particular may benefit from robotic coaches. Personality in HRI has been surveyed in prior work} \cite{robert2020review}. \hl{Prior work has found that participants with lower Neuroticism and higher Conscientiousness had a higher well-being increase response to a robotic coach} \cite{jeong2020robotic}. Another study exploring a robot coaching mindfulness found \hl{that people with low Conscientiousness gave better ratings for a robot's attributes, and people with low Neuroticism enjoyed a robotic coach more} \cite{bodala2020creating}. However, \hl{prior work has not yet explored the impact of the personality trait Introversion on participants' perceptions of robotic coaching.} Future research should further investigate whether and how participants' personality traits, especially introversion, might impact their perceptions and experience of a robotic coach. \textcolor{black}{Additionally, future research should examine how a robot's personality can be designed to best suit a coaching interaction, as prior work has shown that e.g., matching a robot's personality to a user's can lead to improved task performance for the user \cite{tapus2008user}. As a first step towards adopting this recommendation, in a follow up work we have examined the co-design of robot personality, as well as the contribution of robot form to perceived robot personality \cite{spitale2023robotic}.}

\subsection{Robot Affective Adaptation}
\label{sec:user_percp_conditions}


Participants remarked both positively and negatively on every level of adaptation (see Table~\ref{table:conditions}). They especially highlighted instances in which the robot made mistakes with regards to empathetic utterances, during affective adaptation (A) or personalised affective adaptation (P). Here, participants sometimes preferred no adaptation (N), in which the robot didn't make affective mistakes. This occasional preference for \textit{neutrality} is in-line with the comments made by the consulting psychologist, who remarked that neutral statements may be preferred by some individuals as the \ac{RC} making mistakes and saying things that do not align with participants' expectations may become jarring (as seen under the themes in Table \ref{table:capabilities} \emph{Responsiveness} and \emph{Adaptation}). 

Yet, some participants remarked being particularly pleased when the robot responded accurately, as seen in Table \ref{table:capabilities} with regards to \emph{responsiveness}. This supports the notion that \textit{continual personalisation} may become even more relevant for long-term interactions, in the form of \emph{active listening}, which would help them engage more with the robot. In particular, participants would hope to see the robot using ``keywords'' from their speech to signify that it understood what had been said, and provide acknowledgement via gesturing and intermittent back-channelling (such as ``Huh'' and ``Okay'' noises) during the participants' self-disclosure.


\noindent \textbf{Insights for Future Design of Robotic Coaches.} \hl{In terms of qualitative data, we found no overwhelming preference for any of the levels of affective adaptation (None, Adaptation, or Personalised adaptation), over the three interaction rounds despite a proclivity for personalised adaptation in the initial interactions}~\cite{Churamani2022CL4HRI}. \hl{This may be due to the robot occasionally making mistakes in the adaptation and personalisation conditions, resulting in both heightened positive (feeling of ``being listened to'' when the robot adapted correctly) and negative experiences (being disappointed when the robot adapted incorrectly) for the participants. Prior research shows that emotional adaptation can increase robot helpfulness} \cite{kuhnlenz2013increasing}, \hl{but that robot adaptation can sometimes have negative impacts for interactions} \cite{kennedy2015robot}. \hl{Future research should focus on comparing both adaptive and non-adaptive interactions with a robotic coach \textit{longitudinally}, to better understand how participants perceive adaptive interactions in the long term. Additionally, as techniques for longitudinal adaptation are developed and become more accurate, the potential for errors may be reduced. As such, affective adaptation may become more useful in the future.} \textcolor{black}{Our recent research efforts are aimed towards improving the understanding on how such an affective adaptation can be achieved in robots, particularly with respect to `critically analysing and reflecting upon what does and does not work and why?'~\cite{Gunes2023ACHRI}.}

\begin{table}[t]

\small
\centering
\caption{Quotes from participants ($Pi, (i=1,2,...20)$) regarding the study conditions. $+/-$ signs indicate positive and negative statements, neutral statements do not contain signs.}
\begin{center}
\begin{tabular}{|C|L|}
\Xhline{0.5 pt}
\rowcolor{gray!40}
\textit{\textbf{Levels of affective adaptation}} &
\textit{\textbf{Quotes from paticipants }}\\
\Xhline{0.5 pt}

\rowcolor{gray!15}
None \textbf{N} &
{P4+ (\textit{Present}): ``I think the robot was performing really well on the present (N) task.’’

P6- (\textit{Present}): ``In general, I think the past (P) was fine, the future (A) was great, and \emph{the present (N) was really bad}.''

P21- (\textit{Past}): ``It wasn't exactly rude in the first one (N), but the first one was like [less friendly].''

P1- (\textit{Past}): ``At the very beginning (N) I responded and it didn't get it, but after that it was very accurate.''

P5+ (\textit{Present}: ``In the second one (N) it was better at responding.''}
\\
\Xhline{0.5 pt}

\rowcolor{gray!30}
Adaptive \textbf{A} &

P17+ (\textit{Past}): ``Except the last round (A) [...], it felt like mechanical sort of questions that came one upon the other. [...] In last round, I felt it did pick up on some things and it understood what’s a good event.''

P3- (\textit{Present}): ``If I said some negative things it still didn't react to the negativity. That made [it] feel a bit mechanical.''

P2+ (\textit{Present)}: ``I think in the second round (A) I was talking about something positive and it said `I'm very sorry, that sounds like a difficult time'.''

P4- (\textit{Future}): ``The robot completely misunderstood everything I said in the last bit (A).'' 

P3- (\textit{Present}): ``So I actually said one negative thing and it said `that's good', so it felt a bit mechanical in that case.''
 \\
\Xhline{0.5 pt}

\rowcolor{gray!15}
Personalized \textbf{P} &
{P3+ (\textit{Future}): ``‘I think the last one (P) was smoother. It was telling me `OK, I hope you achieve these things'. The last one was easier. It felt more meaningful but I don’t know. Maybe I got used to it.''


P10- (\textit{Future}): ``I don't think it understood me [in the last session] (P).''

P5- (\textit{Future}): ``Sometimes it did really well and sometimes it didn't and I can't really tell why, maybe I was just talking too fast or saying different things. [...] in this third one (P) it just gave the same response.''

P6 (\textit{Past}): ``With the past (P) it was okay, but I feel like it was sort of expecting me to give answers that were more dramatic than the ones I gave. It was saying things like, `Oh, that sounds tough', but I wasn't necessarily feeling like the things were really horrible.''
}\\
\Xhline{0.5 pt}

\end{tabular}
\label{table:conditions}
\end{center}
\end{table}

\subsection{Novelty and Habituation}

\begin{table}[t]
\small
\centering
  \caption{Quotes from participants ($Pi, (i=1,2,...20)$) regarding the phenomena of Novelty and Habituation, and Suspension of Disbelief during RC interaction. $+/-$ signs indicate positive and negative statements, neutral statements do not contain signs.}
  \label{table:phenomena}
  {
  \begin{tabular}{|C|L|}\Xhline{1.0 pt}   
    \rowcolor{gray!40}
    \textit{\textbf{Phenomenon}} & \textit{\textbf{Quotes from participants}}\\ 
    \Xhline{1.0 pt}
    
    \rowcolor{gray!15}
    Novelty and habituation
    & 
    \textbf{Exercises}
    
    P17: ``I felt the last round, [...] I got more used to the pattern and that is probably the bias here, but it might have also been that it understood things more clearly in the last round.’’
    
     P8: ``I think it was very useful, but after the first run, I could predict what exactly was going to happen, and what questions I would be asked, so I felt more prepared. But consequentially, that meant I wasn't thinking or reflecting enough for that.'' 

    \textbf{Robot}
    
    P11: ``[...] the surprise, it was weird to be talking to a robot. It's the surprise when it first talked and moved, and the first time it would say certain things, that were kind of funny as well. 
    I think once I got used to it, and it didn't take very long to get used to it, the weirdness wore off early. [...] Then I was like, `Oh OK, cool that's what it's trained to do', but it caught me off guard.’’

    P20: ``
    Initially I was a bit creeped out by the smile, but it's actually fine. 
    It seemed completely fine, maybe a bit surprising at first when there's some weird arm movement. But otherwise easy to get used to. And even though it was weird at the beginning, I got very used to the appearance after just a few minutes.’’
    
    P11: ``At the beginning I kept laughing and [thinking], this is not the purpose of [the interaction, if] I'm just sitting here laughing [...]. After I got over laughing I started paying closer attention.’’
    \\
    \Xhline{0.5 pt}
    \rowcolor{gray!40}
    Suspension of disbelief
    & P2: ``The fact that this is a study might have skewed my response towards the robot [...]. 
    The experience, especially it being in such a controlled environment [is] a bit strange [...] I felt a lot more observed. I think that definitely did play into how open I was with responses.'' 
    
    P2: ``I think initially it felt a bit strange to be talking to a robot, but I think it really did get a lot better through the exercises. It felt more natural somehow, once you get over the initial strangeness of the whole thing. [...] The entire concept, [...] talking to an inanimate object about your feelings, and your thought process. 
    [...] So it feels like 'am I just saying this, am I just shouting things out into the void'? Or is it actively responding to me?’’
    
    
    P9: ``It was quite strange because you don't really know how much the robot is reacting to what you're actually saying, or to keywords. [...] Even though you have someone or something there to listen to you, it's actually not listening to you. [...] You start questioning, 'what type of interaction am I having'? But it does help to talk about it, even if it is out loud.’’
    
    
    

    P4: ``I know the domain so I was conscious. I was trying to reverse engineer how Pepper works.''
    
    P9+: ``It's quite well made in terms of creating the environments that you might need for a coaching session.’’
    \\
    \Xhline{1.0 pt}
  \end{tabular}
  }

\end{table}


In addition to the themes outlined above, the phenomena of \emph{novelty and habituation}, and \emph{suspension of disbelief} were evident in the analysis of qualitative data (Table \ref{table:phenomena}). 

Participants described their experience with \emph{novelty and habituation} \cite{hoffman2020primer}, both to the exercises and the robot itself. Participants noted that their experience changed throughout the sessions. Habituation was seen as a positive by P17, who understood the exercises more clearly throughout the sessions. P8 on the other hand saw this as both a positive and a negative, noting feeling more prepared but reflecting less due to being able to predict the interactions. Participants described feelings of \emph{surprise} toward the RC when first encountering it. However, participants habituated to the robot over time, after ``getting used to it''. 

\noindent \textbf{Insights for Future Design of Robotic Coaches.} These observations suggest that in order to create successful longitudinal (whether longer single sessions, or multiple repeated sessions) interactions with a robot, the robot should follow an interaction pattern to be understandable, but with variations introduced in order to keep the participant engaged in the exercises. \textcolor{black}{Such an interaction could, for instance, use the pre-defined structure of a Positive Psychology exercise, as is done in coaching \cite{seligman2002positive}, and scaffold this exercise with acknowledging, adaptive robot utterances. Other approaches to increased engagement might include enabling the user to independently choose their preferred exercise from a set of options, or the robot proactively suggesting exercises to the user based on their historical preferences or data.} \textcolor{black}{In a study building on this finding, we examined such an approach to robotic interaction design. We scaffolded a pre-defined structure of a Positive Psycholgy exercise with adaptive robot utterances -- that is, we included adaptive utterances between the pre-defined utterances where appropriate \cite{spitale2023vita}. The adaptive utterances were generated using an LLM \cite{spitale2023vita}.}

\subsection{Suspension of Disbelief}

Suspension of disbelief is a concept that originates from the examination of the willing participation in a fiction by an audience \cite{packman1991incredible}. It has been previously defined in HRI \cite{duffy2012suspension} as the ``person's willingness to suspend their disbelief of what is living in physical social robotics'', and that this may affect how the person perceives the robot. Suspension of disbelief is also relevant in well-being practices, as the optimism --- a foundational concept of Positive Psychology --- has been described as a ``willing suspension of disbelief'' \cite{melnick2005willing}.

While participants did not explicitly describe their experiences as a suspension of disbelief, but described feelings of strangeness and being conscious of how they responded to the robot (Table \ref{table:phenomena}). For example, P2 said that being aware of the robot not actually understanding what they were saying made them less engaged. P2 described talking to an inanimate object as ``shouting into the void'', as they weren't sure what it was actually comprehending. 

Participants also noted that the robot's form contributed to their lack of suspension of disbelief, citing its hands as too human. P4, who has more extensive experience with robots than the majority of participants, noted that they were trying to ``reverse engineer'' the coach's functioning as they were speaking to it. These observations suggest that there is a delicate balance to be struck with participants being able to willingly suspend their disbelief in order to benefit from the coaching, but also understand its level of functioning in order to shape their expectations. 

\noindent \textbf{Insights for Future Design of Robotic Coaches.} In the future, participants' finding it challenging to suspend their disbelief could be mitigated by informing on the capabilities of the RC beforehand. Giving participants sufficient information of what they can expect from the RC may help them better focus on the exercise, rather than the capabilities of the robot. \textcolor{black}{Future research could also examine how participants' trait absorption, that is, their capacity to ``become immersed'' in activities involving new experiences or imagination \cite{wild1995role}, correlates with their propensity for suspension of disbelief. Such investigation might inform how participants' personal traits influence how they may benefit from a robotic coach.}


\section{Conclusion}


\hl{Our study consisted of a single-session user study $(n = 20)$ where participants interacted with a robotic Positive Psychology coach, in a controlled laboratory environment. We based our user-centred design of the robotic coach on our prior research} \cite{axelsson2021roman}, \hl{and iterated upon our design, and particularly the robotic coach's utterances, together with a professional well-being coach. This study is part of a larger effort to iteratively design a robotic well-being coach.}

\hl{We found that participants perceived the robotic coaching positively. They found the coaching particularly helpful in that the robot asked the correct questions to help them think of new positive experiences in their life. In future robotic coach design, some features and capabilities of the robotic coach should be improved. In particular, participants wanted the robotic coach's utterance content to be more responsive, acknowledging what they said, and helping them ``feel listened to''. The robot also made conversational mistakes such as interrupting the users. In the future, the robot's turn-taking capabilities should be improved, and repair strategies for robot mistakes should be explored. Additionally, future research should explore whether robotic coaches could be particularly helpful in circumstances in which users may feel isolated. Users mentioned that these circumstances might include a hospital stay, where patients have limited social interaction, or a pandemic where social distancing guidelines are followed. Additionally, future research should explore how participants with different personality traits, especially introversion, could benefit from robotic coaches.}

\hl{In terms of affective adaptation, no clear preference for no adaptation, non-personalised adaptation, or personalised adaptation was found in the qualitative data. Participants remarked both positively and negatively on all three levels of affective adaptation (no adaptation, adaptation, and personalised adaptation). Some participants had positive perceptions of the adaptation, noting that it helped them feel listened to. However, since the affective adaptation sometimes resulted in the robot making a mistake (that is, adapting incorrectly to the participants' affect), this resulted in negative perceptions of this conditions. In the future, affective adaptation should be explored in longitudinal interactions, which may help improve its accuracy through increased interactions with users, especially in the case of personalised adaptation where the accuracy is continually improved. In particular, a between-subjects study comparing personalised affective adaptation and no affective adaptation could shed light on this design feature. The results of our study in this regard were limited by the within-subjects comparison of the levels of adaptation (that is, all participants experienced all levels of adaptation), and the single-session nature of the interaction. \textcolor{black}{Additionally, order effects could have affected the results. Although the condition orders were counterbalanced, there was some imbalance across this counterbalancing due to participants dropping out after the condition orders had already been allocated.} Future work should explore the design feature of adaptation and personalisation in longitudinal robotic coaching interactions.}

\vspace{5mm}

\textbf{\large{Acknowledgements}}

We would like to thank Katherine Parkin for her valuable help with Positive Psychology exercises. We are thankful to all the study participants.

\noindent\textbf{Funding:} M.~Axelsson was funded by the Osk. Huttunen foundation and the EPSRC under grant EP/T517847/1. N.~Churamani's work was funded by the EPSRC under grant EP/R513180/1 (ref. 2107412). H.~Gunes is supported by the EPSRC under grant ref. EP/R030782/1. A.~Çaldır contributed to this work while undertaking a summer research study at the Department of Computer Science and Technology, University of Cambridge.

\noindent\textbf{Data Access Statement:} Raw data related to this publication cannot be openly released as the raw data contains videos and transcripts of the participants’ interaction with the robot, which were impossible to anonymise.

\begin{acronym}
    \acro{AC}{Affective Computing}
    \acro{AI}{Artificial Intelligence}
    \acro{ASR}{Automatic Speech Recognition}
    \acro{AU}{Action Unit}
    \acro{BMU}{Best Matching Unit}
    \acro{CL}{Continual Learning}
    \acro{CLIFER}{Continual Learning with Imagination for Facial Expression Recognition}
    \acro{CNN}{Convolutional Neural Network}
    \acro{EDA}{Electrodermal Activity}
    \acro{FACS}{Facial Action Coding System}
    \acro{FER}{Facial Expression Recognition}
    \acro{FSM}{Finite-state Machine}
    \acro{FoI}{Frame-of-Interest}
    \acro{GAD7}{General Anxiety Disorder}
    \acro{GDM}{Growing Dual-Memory}
    \acro{GWR}{Growing When Required}
    \acro{HRI}{Human-Robot Interaction}
    \acro{MFCC}{Mel-Frequency Cepstral Coefficients}
    \acro{ML}{Machine Learning}
    \acro{MLP}{Multilayer Perceptron}
    \acro{NLG}{Natural Language Generation}
    \acro{NLP}{Natural Language Processing}
    \acro{LSTM}{Long Short-Term Memory}
    \acro{PANAS}{Positive and Negative Affect Schedule}
    \acro{PD}{Participatory Design}
    \acro{PHQ9}{Participant Health Questionnaire}
    \acro{PP}{Positive Psychology}
    \acro{RC}{Robotic Coach}
    \acro{RL}{Reinforcement Learning}
    \acro{ReLU}{Rectified Linear Unit}
    \acro{ROS}{Robot Operating System}
    \acro{RoSAS}{Robotic Social Attributes Scale}
    \acro{SAR}{Socially Assistive Robot}
    \acro{TA}{Thematic Analysis}
    \acro{TTS}{Text-to-Speech}
    \acro{WEMWBS}{Warwick-Edinburgh Mental Well-being Scale}
    \acro{WoZ}{Wizard of Oz}
\end{acronym}

\bibliographystyle{plainnat}
\bibliography{main}

\newpage
\appendix

\section{Post-interaction interview}\label{secA1}

The questions asked during the semi-structured post-interaction interview:

1. Robot
  \begin{itemize}
    \item What do you think was good about the robot’s appearance? What did you like?
    \item What do you think was bad about the robot’s appearance? What would you change?
    \item How appropriate do you think the robot was for helping you focus on the positive aspects in your life? How appropriate do you think the robot was for helping you foster a positive attitude toward things that happen in your life?
    \item How appropriate do you think the robot was for these coaching tasks?  
    \item How useful do you think the robot was for these coaching tasks? 
    \item How beneficial do you think the robot was for these coaching tasks?
  \end{itemize}
  
2. Behaviour
  \begin{itemize}
    \item What do you think was good about the robot’s behaviour? What did you like?
    \item What do you think was bad about the robot’s behaviour? What would you change?
    \item How appropriate do you think the robot’s behaviour was for these coaching tasks?  
    \item How useful do you think the robot’s behaviour was for these coaching tasks?  
    \item How beneficial do you think the robot’s behaviour was for these coaching tasks?  
    \item Do you think the robot understood what you said?
    \item Do you think the robot understood how you felt?
    \item Do you think the root adapted to what you said and did?
    \item Did you notice any differences?     
  \end{itemize}

\end{document}